\documentclass[letterpaper, 10pt]{article}

\usepackage[margin=1in, letterpaper]{geometry}

\usepackage[T1]{fontenc}
\usepackage{stix2}

\usepackage{graphicx}
\usepackage{booktabs}
\usepackage{array}
\usepackage{tabularx}
\usepackage{caption}
\usepackage{float}
\usepackage{hanging}
\usepackage{setspace}
\usepackage{titlesec}
\usepackage{fancyhdr}
\usepackage{lastpage}
\usepackage{enumitem}
\usepackage{hyperref}
\usepackage{xcolor}
\usepackage{hyperref}
\usepackage{xurl}

\titleformat{\section}
  {\fontsize{13}{15.6}\bfseries}
  {}
  {0pt}
  {}
\titlespacing{\section}{0pt}{12pt}{4pt}

\titleformat{\subsection}
  {\fontsize{11}{13.2}\bfseries\itshape}
  {}
  {0pt}
  {}
\titlespacing{\subsection}{0pt}{10pt}{4pt}

\titleformat{\subsubsection}
  {\fontsize{10}{12}\bfseries}
  {}
  {0pt}
  {}
\titlespacing{\subsubsection}{0pt}{8pt}{4pt}

\hypersetup{
  colorlinks=false,
  pdfborder={0 0 0}
}

\begin{document}

\begin{center}
{\fontsize{20}{24}\selectfont\bfseries How Verification Mechanisms Alter Cultural\\Signals in Employer Reviews}

\vspace{1pt}
{\fontsize{12}{14}\selectfont\itshape Full Paper}

\vspace{12pt}
\setlength{\tabcolsep}{6pt}
\renewcommand{\arraystretch}{1.5}
\begin{tabular}{>{\centering\arraybackslash}p{0.45\textwidth}>{\centering\arraybackslash}p{0.45\textwidth}}
\parbox[c][1.8cm][c]{0.43\textwidth}{\centering
{\fontsize{13}{15}\selectfont\textbf{Vladimir Martirosyan}}\\
{\fontsize{13}{15}\selectfont University of Maryland}\\
{\fontsize{13}{15}\selectfont martiros@umd.edu}}
&
\parbox[c][1.8cm][c]{0.43\textwidth}{\centering
{\fontsize{13}{15}\selectfont\textbf{Rachit Kamdar}}\\
{\fontsize{13}{15}\selectfont University of Maryland}\\
{\fontsize{13}{15}\selectfont rkamdar@umd.edu}}
\end{tabular}
\end{center}
\vspace{6pt}

\section{Abstract}

Online reviews shape impressions across products and workplaces, and employer reviews in particular combine narratives and ratings that reflect organizational culture. Two major platforms illustrate contrasting approaches to reviewer credibility: Glassdoor permits fully anonymous posts, while Blind requires employment verification while preserving anonymity. We ask how verification changes reviews. Evidence suggests verified reviews can be more trustworthy, yet verification can also erode authenticity when expectations are unmet. We use the Competing Values Framework (clan, adhocracy, hierarchy, market) and the CultureBERT model developed by Koch and Pasch (2023) to analyze over 300k ratings. We find that Blind reviews emphasize clan and hierarchy while Glassdoor skews positive and highlights clan and market. Verification alone does not remove bias but shifts how culture is represented. Job seekers using different platforms receive systematically different signals about workplace culture, which affects application decisions and job-matching.

\vspace{6pt}
\textbf{Keywords}

Online Reviews, Digital Platforms, Natural Language Processing

\section{Introduction}

While similar to consumer review sites, employer platforms differ by centering on current or former employees who evaluate management quality and work culture (Pavithra and Westbrook, 2022). Because reviews often raise sensitive issues, anonymity is standard (Cloos, 2021); many platforms permit fully anonymous posts, often in exchange for access to other users' reviews. Some platforms go further by requiring corporate email verification before a review can be submitted, though the reviewer's identity remains hidden from readers. This anonymity-friendly design sets company review sites apart from other review forums and encourages frank, insider perspectives on workplaces (Pavithra and Westbrook, 2022).

Companies rated higher on review platforms often obtain proportionally greater average application rates (Sockin and Sojourner, 2023). Consequently, it is in a firm's strategic interest not only to encourage employee reviews, but also to assess which platforms align best with their employer branding goals. At the same time, studies indicate that discrepant or highly polarized reviews, where employee opinions differ sharply, can lower job seekers' intentions to pursue employment at the company (K\"{o}nsgen et al., 2018). This underscores that job seekers are also sensitive to the consistency and credibility of employee sentiment.

If strong ratings draw applicants, then one must ask: Are some firms posting fake reviews to burnish their image? In the realm of consumer reviews, this kind of fraud is well-documented: independent hotels were found to receive significantly better ratings on TripAdvisor (an open site) than on Expedia (which only allows verified customers), suggesting that some hotels posted ``promotional'' fake reviews on the open platform to boost their reputation (Mayzlin et al., 2014). By analogy, it is reasonable to suspect similar behavior in employer reviews. A Wall Street Journal investigation found that several companies orchestrated sharp spikes in five-star Glassdoor ratings by soliciting reviews from enthusiastic employees (Winkler and Fuller, 2019). Quantifying the exact scale of fake reviews on labor platforms is difficult, in part because successful fakes are hard to detect and may be too short for deeper analysis, but the risk is acknowledged. Notably, Glassdoor does not require any proof that a reviewer actually works at the company, anyone with a valid email can sign up and write an employer review, creating an opening for inauthentic entries (Martin-Fuentes et al., 2018). Researchers have pointed out that a system which ``allows anonymous users to give opinions about any establishment without [verification]'' faces a credibility threat in that unscrupulous parties can exploit it (Martin-Fuentes et al., 2018).

To uphold content integrity, many platforms use reviewer verification to confirm that posts come from actual employees. Unlike e-commerce sites that only flag ``Verified Purchases,'' job review platforms can require company email or employment confirmation. Blind's model is a case in point: it only allows verified employees (via their work email domains) to contribute content, thereby dramatically reducing the likelihood of outsiders or bots posting fake company reviews (Chaudhary et al., 2023). These considerations motivate two research questions:

\begin{enumerate}[leftmargin=*, label=\arabic*.]
\item How do rating distributions differ by platform verification status?
\item How does verification shape the cultural signals embedded in employee reviews?
\end{enumerate}

Despite increasing use of verification mechanisms in job review platforms, little academic work has examined how these features shape the reviews themselves, whether in terms of rating distribution, perceived credibility, or the representation of workplace culture. We address this gap by comparing verified and non-verified review platforms through the lens of cultural dimensions, offering new insight into how verification may influence both the tone and substance of job-related feedback.

\section{Literature Review}

Online job platforms expose two critical metrics: the rating distribution of employee reviews (often taken as a proxy for employee satisfaction (H\"{o}llig, 2021) and qualitative cultural signals (Pacelli et al., 2022). These are vital because employee satisfaction levels as reflected in rating distributions are closely tied to organizational outcomes like lower turnover, higher productivity, and even profitability (Ding et al., 2025). Likewise, workplace culture heavily influences satisfaction; firms with supportive, team-oriented cultures generally see more satisfied employees and thus better reviews (Ding et al., 2025).

The Competing Values Framework (CVF) provides a useful lens on culture, categorizing types (e.g. clan, adhocracy, market, hierarchy) and their effects. For example, collaborative ``clan'' cultures are associated with significantly higher job satisfaction (Hartnell et al., 2011), often manifesting in more positive employee ratings. Prior research applying the CVF to Glassdoor reviews found that stronger cultural satisfaction in employee feedback corresponded to higher employee referral intention (willingness to recommend the company; Seo and Lee, 2021), underlining that culture signals in reviews have tangible importance for companies' talent attraction and reputation.

Introducing verification of reviewers may be associated with differences in both rating distributions and cultural signals, though the direction of these associations remains unclear. On one hand, anonymity has been associated with inflated positive reviews as well as extreme negative outliers (Deng et al., 2021), suggesting that verification may correspond with more balanced and credible assessments. On the other hand, requiring verification could amplify certain voices or introduce new forms of bias, potentially skewing perceptions in a different way (Mardumyan and Siret, 2023). Whether verification corresponds with more representative cultural signals or simply relates to different manifestations of existing biases is an open question this study aims to explore using comparative analysis across platforms.

A key interpretive challenge is disentangling verification effects from other structural platform differences. Glassdoor and Blind differ not only in verification requirements but also in audience composition, industry focus, and interaction model, all of which independently shape review content (Cloos, 2021; Martin-Fuentes et al., 2018). Observed differences may therefore reflect reviewer self-selection or platform culture rather than verification per se. We address this by restricting comparisons to companies present on both platforms and conducting a within-firm analysis using Amazon reviews, where company and industry are held constant.

\section{Data and Methods}

We leveraged two large datasets of employee reviews. Blind's reviews were obtained from an open dataset scraped from the TeamBlind platform (covering 25+ major tech and consulting companies). Blind is an anonymous professional forum where ``work email-verified professionals'' share reviews about their workplace, including overall ratings, pros, cons, and other commentary. The collected Blind dataset spans the platform's launch up to May 2022, comprising tens of thousands of reviews (e.g. \textasciitilde9,903 reviews for Amazon alone). For Glassdoor, we used the publicly available ``Glassdoor Job Reviews'' dataset, which contains approximately 850,000 reviews from Glassdoor users across many companies and years. Each Glassdoor entry similarly provides an overall rating, date, employer name, and free-text fields for ``Pros'' and ``Cons.''

From both sources, we extracted each review's company name, date, and numeric rating (on a 1--5 scale). We created a textual corpus of reviews by concatenating the ``Pros'' and ``Cons'' fields for each entry into a single document per review (this combined text captures the reviewer's overall commentary inputted in both fields). Data cleaning steps included removing duplicates, standardizing company names, and parsing the Blind JSON data (which was nested by company) into a flat table of reviews. Company URLs were used to identify company names in the Glassdoor dataset.

Our quantitative analysis has two parts. First, we examine rating distributions on each platform. We computed summary statistics such as the mean and standard deviation of ratings for Glassdoor vs. Blind, and visualized the distributions (e.g. via histograms) to assess skewness and variance. To test whether observed differences were statistically significant, we employed two-sample t-tests for mean rating comparisons and two-proportion z-tests for share-level comparisons across rating categories. Effect sizes were assessed using Cohen's \textit{d} to distinguish statistical from practical significance. This allowed us to compare overall rating tendencies, whether one platform skews more positive or exhibits greater volatility. Second, we analyzed organizational culture signals present in the review text. We applied the pre-trained CultureBERT language model (Koch and Pasch, 2023) to each review's text to classify it into one of the four culture types defined by the Competing Values Framework: Clan, Adhocracy, Market, or Hierarchy. CultureBERT is a transformer-based model fine-tuned on employee reviews to detect dominant culture traits in text (e.g. whether a review's content aligns more with a ``family-like, mentoring'' Clan culture or a ``competitive, results-oriented'' Market culture, and so on). For each platform, we aggregated the predicted culture labels to see which culture dimensions are most frequently reflected in the reviews. This textual analysis complements the numerical ratings, letting us gauge if verification status influences not just how high or low employees rate their company, but also what aspects of culture they emphasize in their comments and how these differ by company.

It is worth situating the two platforms as comparable yet meaningfully distinct environments. Although both Glassdoor and Blind allow anonymous employee reviews and cover compensation and workplace culture, they differ in several structural dimensions that extend beyond verification alone, including their target industries and access models. Table 1 summarizes these key similarities and differences. We acknowledge that these platform-level heterogeneities represent potential confounding factors and our analysis is careful to treat verification as a focal mechanism while remaining attentive to alternative explanations. By making these differences explicit, we aim to ground our comparative claims in the full context of each platform's design rather than attributing observed differences to verification in isolation.

To assess the validity of CultureBERT's classifications, we conducted a manual coding exercise on a random sample of 20 reviews to one of the four CVF culture types, following the labeling approach of Koch and Pasch (2023) and guided by the culture dimension descriptions from Cameron and Quinn (2011). Each review was assigned to the dimension that best fit the overall tone, consistent with the assumption that one culture type typically dominates (Cameron and Quinn, 2011). Model predictions agreed with at least one coder on 65\% of reviews, consistent with the validation benchmarks reported by Koch and Pasch (2023) for transformer-based culture classification on employee text. Accordingly, we treat CultureBERT outputs as probabilistic signals of cultural emphasis rather than definitive classifications, and our inferences focus on distributional patterns across large samples rather than individual review-level predictions.

\begin{table}[H]
\begin{tabular}{|p{3cm}|p{5cm}|p{5cm}|}
\hline
\textbf{Feature} & \textbf{Glassdoor} & \textbf{Blind} \\
\hline
Anonymity & Users share reviews and salaries anonymously. & Users share reviews and salaries anonymously. \\
\hline
Company Ratings & Provides star ratings for culture, CEO, and benefits.&  Employees rate their companies.\\
\hline
Salary Data & Structured database of salary ranges by role.&  Real-time crowdsourced ``TC'' (Total Comp) threads.\\
\hline
Verification Access & Anyone can read reviews after posting one.&  Full access requires a current work email.\\
\hline
Interaction Style & Primarily a collection of one-way reviews.&  A Reddit-style forum for active Q\&A.\\
\hline
Main Audience & All industries & Primarily tech industries \\
\hline
\end{tabular}
\centering
\caption{Key Platform Characteristics}
\end{table}
As Table 1 illustrates, the most salient structural distinction remains the verification mechanism — Blind requires a valid corporate email to post, while Glassdoor does not. Although Blind's primary audience skews toward tech professionals while Glassdoor spans all industries, our analysis deliberately focuses on the subset of companies present in both platforms, which are predominantly major tech and consulting firms. This overlap in firm coverage substantially narrows the audience gap between platforms. Verification thus remains the primary lever of interest, even as we remain attentive to residual differences in platform culture and behavior.

\section{Results}

\subsection{Rating Distribution Across Platforms}

\begin{figure}[H]
\centering
\includegraphics[width=.8\textwidth]{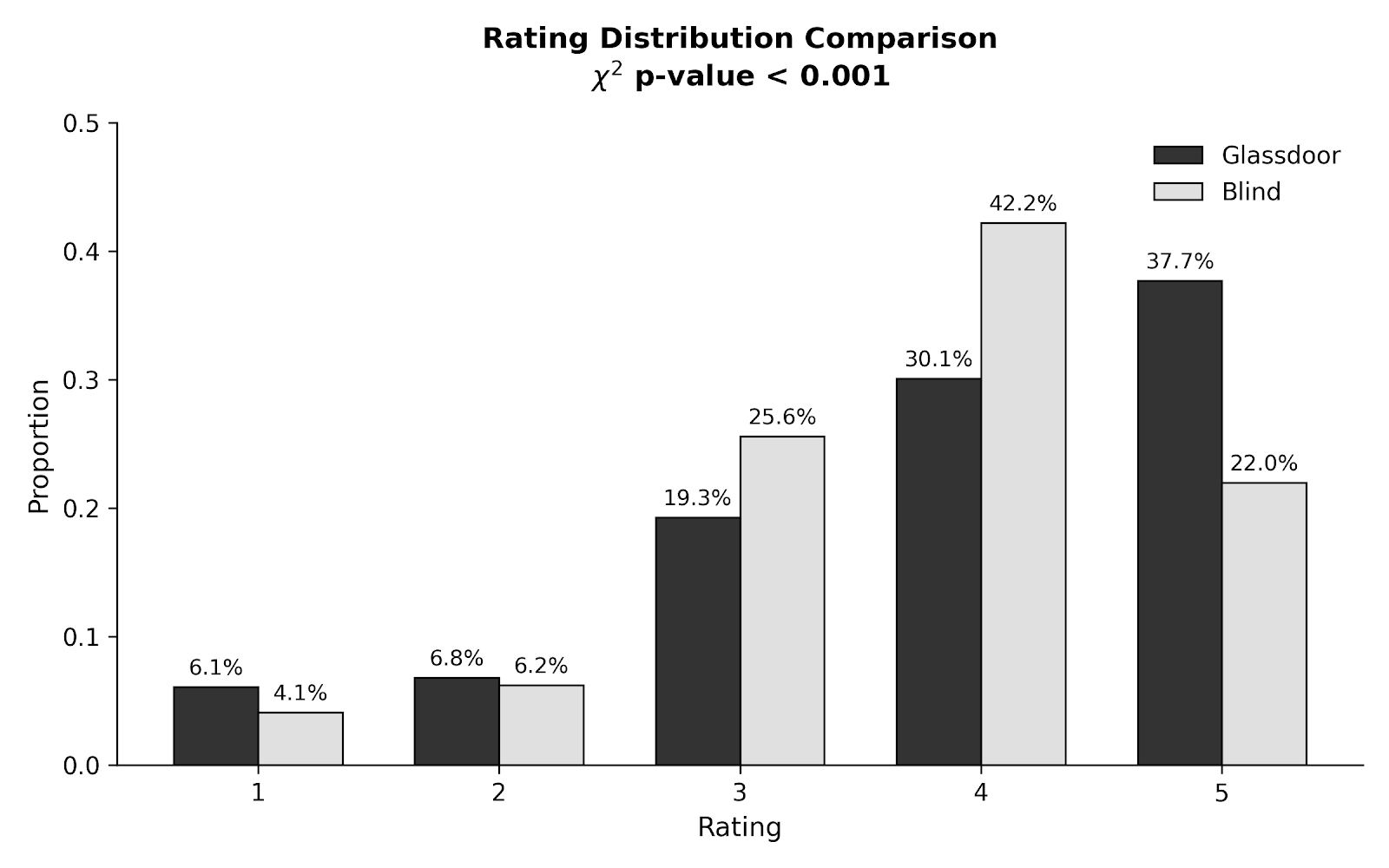}
\caption{Rating distribution for both Glassdoor and Blind}
\end{figure}

We first compare the overall rating distributions in review descriptions 
between Glassdoor and Blind. As shown in Figure 1, the distributions differ 
significantly across the two platforms ($\chi^{2}$, p $<$ 0.001). Glassdoor 
reviews display a clear positivity bias: 37.7 percent of reviews award the 
maximum five stars, compared to only 22.0 percent on Blind. By contrast, 
Blind contains a substantially higher share of moderate evaluations, with 
42.2 percent of reviews at four stars (versus 30.1 percent on Glassdoor) and 
25.6 percent at three stars (versus 19.3 percent on Glassdoor). Notably, 
Glassdoor also exhibits slightly higher shares of negative reviews, with 
6.1 percent one-star and 6.8 percent two-star reviews compared to 4.1 percent 
and 6.2 percent on Blind respectively, suggesting that unverified platforms 
polarize sentiment in both directions. The mean rating difference of 0.15 
points is statistically significant (t = 20.36, p $<$ 0.001), though the 
modest effect size (Cohen's $d$ = 0.13) suggests that verification reshapes 
the \textit{distribution} of ratings more than it shifts their average level. 
Specifically, verification appears to compress extremes on both ends, 
redistributing mass toward the moderate range, consistent with more calibrated 
assessments from verified reviewers.

To isolate platform effects from compositional differences, we conduct a within-firm analysis using Amazon reviews. We conduct a more detailed analysis using Amazon reviews, 
which constitute our largest single-company sample (N = 108,521). As shown 
in Table 3, the within-company results largely reinforce the aggregate 
findings but reveal one important divergence: unlike the aggregate pattern, 
Amazon's Blind reviewers leave \textit{more} negative reviews than their 
Glassdoor counterparts, with 10.4 percent two-star and 8.1 percent one-star 
reviews on Blind versus 7.8 percent and 7.2 percent on Glassdoor. This 
suggests that for a high-pressure firm like Amazon, verification may embolden 
frank negative feedback from employees who feel more secure expressing 
dissatisfaction through a credentialed platform. The five-star share 
difference nonetheless remains large at 25.0 percentage points, and the 
overall mean gap widens to 0.47 points (p $<$ 0.001), reinforcing that the 
dominant effect of verification is suppression of extreme positivity.

\begin{table}[H]
\centering
\begin{tabular}{|l|c|c|c|}
\hline
 & \textbf{Glassdoor} & \textbf{Blind} & \textbf{Difference} \\
\hline
Mean Rating & 3.86 & 3.72 & 0.15*** \\
\hline
5-star Reviews (\%) & 37.7 & 22.0 & 15.7 pp*** \\
\hline
4-star Reviews (\%) & 30.1 & 42.2 & $-$12.1 pp*** \\
\hline
3-star Reviews (\%) & 19.3 & 25.6 & $-$6.3 pp*** \\
\hline
2-star Reviews (\%) & 6.8 & 6.2 & 0.7 pp*** \\
\hline
1-star Reviews (\%) & 6.1 & 4.1 & 2.0 pp*** \\
\hline
High Rating \% & 67.8 & 64.2 & 3.6 pp*** \\
\hline
\end{tabular}
\caption{Platform Effects on Employee Ratings --- All Companies}
\vspace{6pt}
Note: *** p $<$ 0.001. pp = percentage points. N = 304,048 reviews (275,184 Glassdoor, 28,864 Blind). High ratings defined as 4-5 stars on five-point scale. Significance tested via two-sample t-test for mean rating and two-proportion z-tests for share comparisons.
\end{table}

\begin{table}[H]
\centering
\begin{tabular}{|l|c|c|c|}
\hline
 & \textbf{Glassdoor} & \textbf{Blind} & \textbf{Difference} \\
\hline
Mean Rating & 3.78 & 3.31 & 0.47*** \\
\hline
5-star Reviews (\%) & 35.3 & 10.3 & 25.0 pp*** \\
\hline
4-star Reviews (\%) & 29.4 & 37.0 & $-$7.5 pp*** \\
\hline
3-star Reviews (\%) & 20.2 & 34.3 & $-$14.0 pp*** \\
\hline
2-star Reviews (\%) & 7.8 & 10.4 & $-$2.6 pp*** \\
\hline
1-star Reviews (\%) & 7.2 & 8.1 & $-$0.9 pp*** \\
\hline
High Rating \% & 64.8 & 47.2 & 17.5 pp*** \\
\hline
\end{tabular}
\caption{Platform Effects on Amazon Employee Ratings}
\vspace{6pt}
Note: *** p $<$ 0.001. pp = percentage points. N = 108,521 reviews (98,621 Glassdoor, 9,900 Blind). High ratings defined as 4-5 stars on five-point scale. Significance tested via two-sample t-test for mean rating and two-proportion z-tests for share comparisons.
\end{table}

\subsection{Organizational Culture Signals}

The analysis of cultural framing in review descriptions reveals additional differences between platforms (Figure 2). Both Glassdoor and Blind emphasize clan culture most strongly, which reflects collaboration and supportive environments. However, this emphasis is more pronounced on Glassdoor, where 41.5 percent of reviews highlight clan attributes, compared to 35.5 percent on Blind. Blind reviews assign more weight to adhocracy (17.8 percent versus 10.2 percent on Glassdoor) and hierarchy (30.4 percent versus 17.9 percent on Glassdoor). Glassdoor reviews, in contrast, more frequently stress market culture (30.3 percent versus 16.2 percent on Blind), showing a stronger focus on external competitiveness and performance.

\begin{figure}[H]
\centering
\includegraphics[width=.9\textwidth]{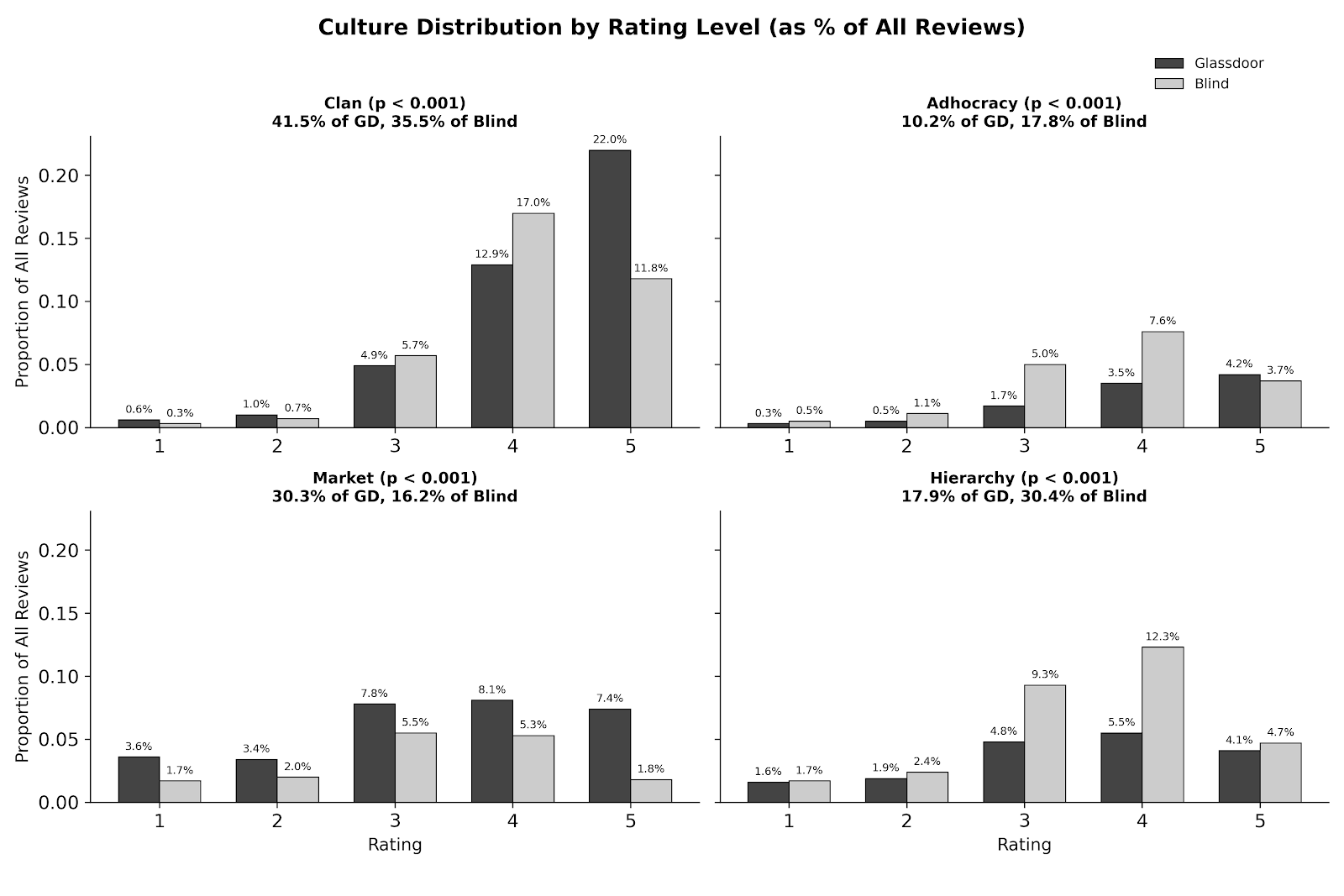}
\caption{Distribution of reviews associated with 4 CVF cultures}
\end{figure}

\subsection{Pros and Cons}

Breaking down reviews into pros and cons provides further insight into how organizational culture types are framed by employees.

\begin{figure}[H]
\centering
\includegraphics[width=.7\textwidth]{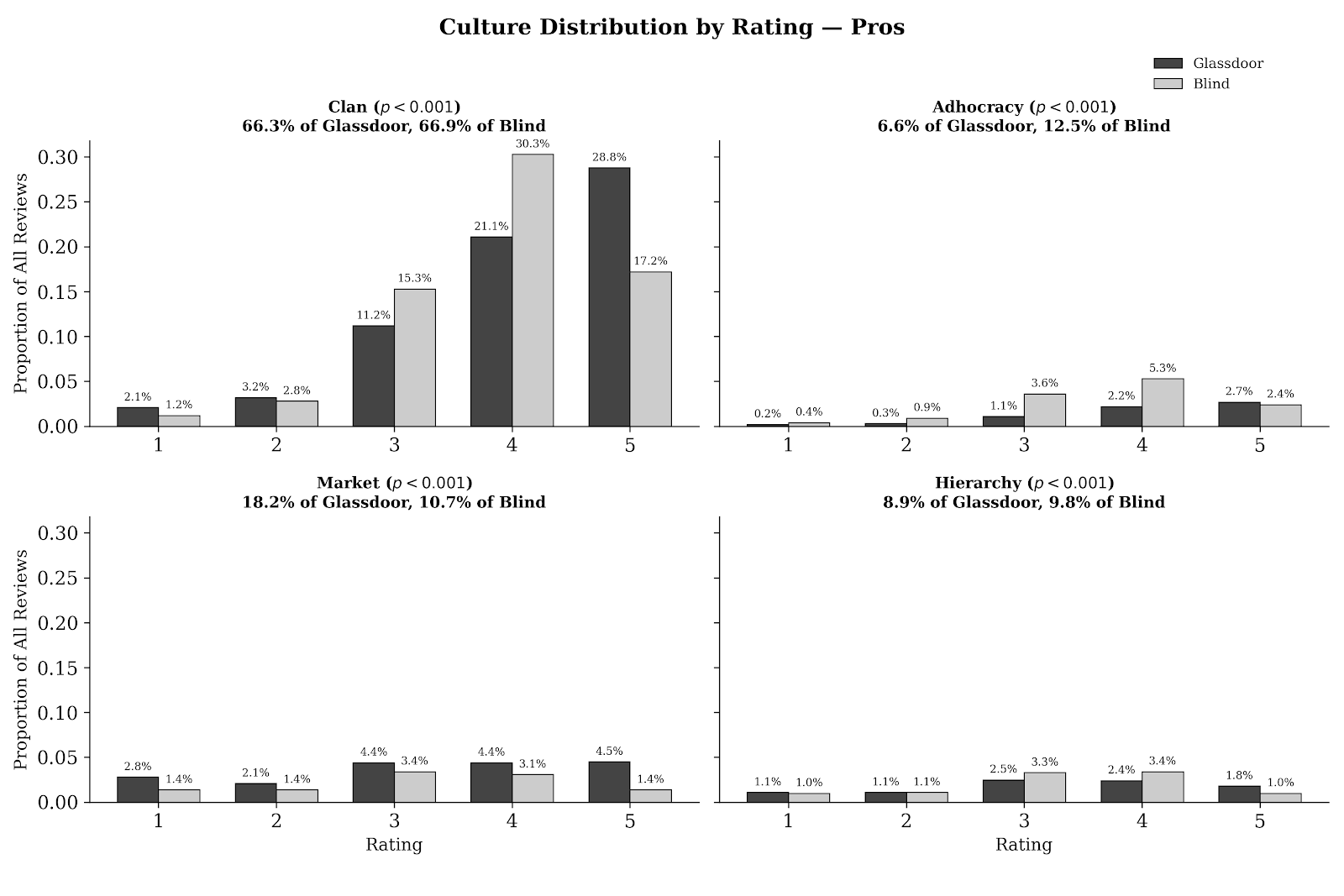}
\caption{Distribution of Pros Section of Reviews Associated with CVF Cultures}
\end{figure}

For pros (Figure 3), both platforms show that clan culture is the most positively framed, with over two-thirds of reviews referencing clan values (66.3 percent on Glassdoor and 66.9 percent on Blind). This suggests that employees tend to highlight supportive, collaborative, and people-oriented environments when describing positive aspects of their workplaces. Adhocracy also appears more in pros on Blind (12.5 percent) compared to Glassdoor (6.6 percent), indicating that verified reviewers may value innovation and adaptability more strongly. Market and hierarchy cultures are much less likely to be emphasized as pros, accounting for under 20 percent combined on each platform.

\begin{figure}[H]
\centering
\includegraphics[width=.7\textwidth]{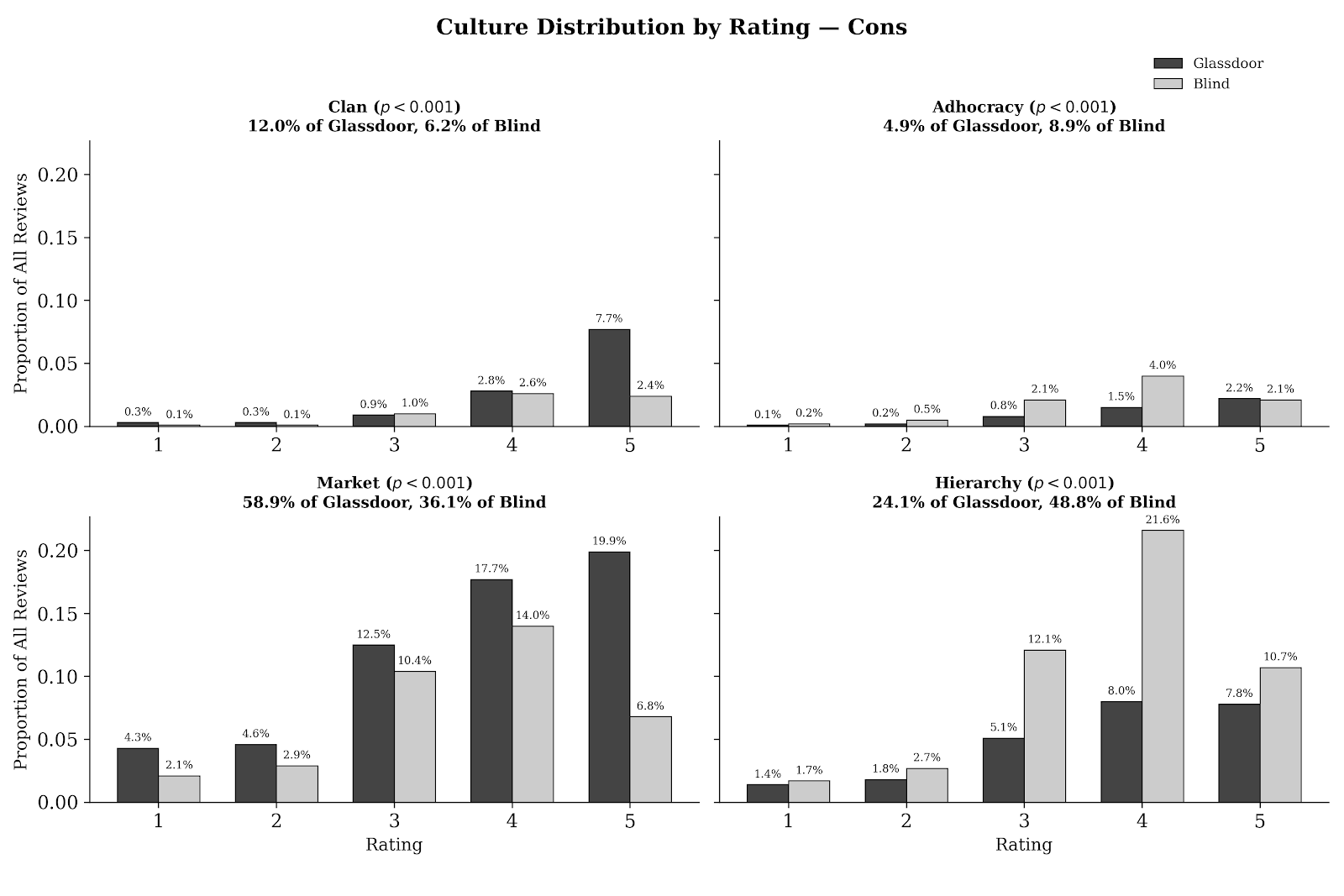}
\caption{Distribution of Cons Section of Reviews Associated with CVF Cultures}
\end{figure}

By contrast, cons (Figure 4) are disproportionately linked to market and hierarchy cultures. On Glassdoor, 58.9 percent of cons highlight market culture, while on Blind 36.1 percent do so. Similarly, hierarchy is mentioned in 24.1 percent of Glassdoor cons and nearly half of Blind cons (48.8 percent). This pattern suggests that control-oriented or competitive environments are more frequently perceived as drawbacks in employer practices. Clan culture appears far less in cons (12.0 percent of Glassdoor and 6.2 percent of Blind), implying that collaborative structures rarely generate negative commentary. Adhocracy is also only weakly associated with cons on both platforms (under 9 percent).

Taken together, these results show a clear polarity: clan culture is predominantly linked with pros, while market and hierarchy cultures are more often framed as cons. Adhocracy plays a more modest role, with Blind reviewers somewhat more likely to describe it positively. This polarity provides evidence that the way employees perceive organizational culture strongly influences whether it is framed as a strength or weakness in their evaluations. Overall, these findings indicate that verification influences both the sentiment of ratings and the types of cultural characteristics emphasized in descriptions. Glassdoor's unverified reviews tend to present organizations more positively and highlight collaboration and competitiveness. Blind's verified reviews, while less inflated, place greater emphasis on hierarchy and innovation, offering a different view of organizational dynamics that may be muted in fully anonymous environments. To assess the generalizability of these platform effects, Table 4 compares regression results across companies. Both Adobe and Amazon demonstrate consistent positive platform effects, with Glassdoor reviews receiving systematically higher ratings than Blind reviews across different organizational contexts.

\begin{table}[H]

\begin{tabular}{|l|c|c|}
\hline
\textbf{Company} & \textbf{Platform Coefficient} & \textbf{Odds Ratio (High Rating)} \\
\hline
Adobe (N = 2,381) & 0.301*** & 1.48*** \\
\hline
Amazon (N = 108,521) & 0.469*** & 2.05*** \\
\hline
Mean Effect & 0.385 & 1.77 \\
\hline
\end{tabular}
\centering
\caption{Comparison of Platform Effects Across Companies}

\vspace{6pt}
Note: *** p $<$ 0.001. Platform coefficient represents increase in rating points for Glassdoor vs Blind. Odds ratios show likelihood of receiving 4-5 star rating on Glassdoor vs Blind. Both companies show consistent positive platform effects.
\end{table}

Amazon results provide compelling validation with exceptional statistical power (N = 108,521). The 0.47-point rating difference and 2.05x odds ratio for high ratings show systematic differences across platforms in rating behavior.

\section{Discussion and Conclusion}

Our analysis shows clear platform effects on ratings (RQ1). Blind's verified reviews are more moderate, while Glassdoor exhibits a higher share of five-star ratings, consistent with polarization in open online reviews (Pavithra and Westbrook, 2022). Although Glassdoor has introduced policies such as give-to-get, positivity remains elevated, likely reflecting impression management and firm-driven encouragement (Figini et al., 2020; Mardumyan and Siret, 2023; Pavithra and Westbrook, 2022). Verification that preserves anonymity appears to support candid but restrained assessments, reducing hyper-positive outliers (Figini et al., 2020; Mardumyan and Siret, 2023). Thus, platform design meaningfully shapes tone.

For culture (RQ2), Blind reviews more often surface hierarchy and adhocracy in cons, whereas Glassdoor highlights clan and market in pros. This pattern aligns with evidence that clan culture is linked to higher satisfaction and favorable perceptions, while hierarchical or sharply competitive environments correlate with negatives (Hartnell et al., 2011; Seo and Lee, 2021). No verification requirement of Glassdoor likely amplifies outward-facing cultural praise, while Blind invites internal critique.

We propose two mechanisms to explain these results. Verified anonymity on Blind fosters an insider audience and perceived safety, encouraging disclosure without fear of retaliation, while still imposing credibility (Figini et al., 2020; Mardumyan and Siret, 2023). Glassdoor's broad, public audience and looser employment verification create incentives for strategic self-presentation and sustained positivity (Mardumyan and Siret, 2023; Pavithra and Westbrook, 2022). Figure 5 synthesizes these mechanisms, illustrating how verification requirements shape reviewer composition and, in turn, the cultural dimensions surfaced in employee reviews, with illustrative company-level examples drawn from our CultureBERT classification results.

\begin{figure}[H]
\centering
\includegraphics[width=\textwidth]{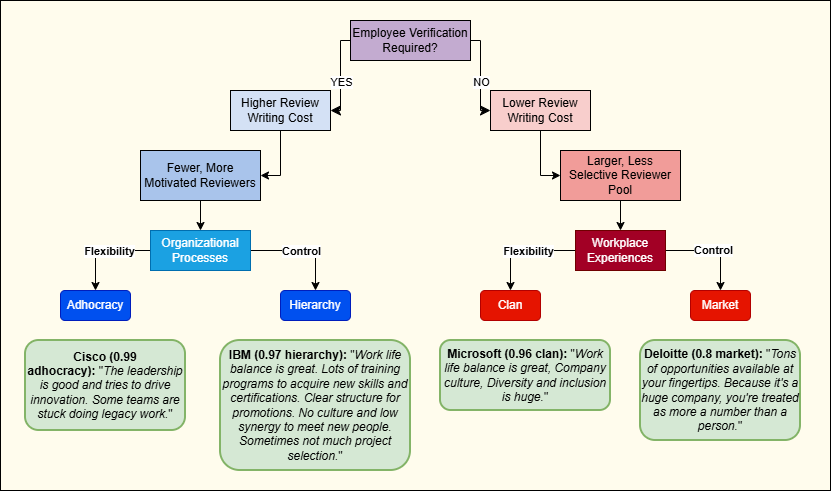}
\caption{Summary Framework: How Verification Shapes Cultural Signals in Employer Reviews}
\end{figure}

Limitations include observational data with different user bases and norms across platforms, potential self-selection, reliance on automated cultural classification with imperfect accuracy, and a cross-sectional design that cannot establish causality (Hartnell et al., 2011; Mardumyan and Siret, 2023). Company-specific shocks during the sample window may also influence comparisons.

Future work can test other platforms and contexts and link cultural signals to outcomes such as turnover or productivity. Qualitative studies could clarify user motives and audience effects observed here (Mardumyan and Siret, 2023; Seo and Lee, 2021).

This study advances an information systems view of employee voice by showing how verification and audience configurations reweigh expressive incentives and costs, thereby modulating both content selection and evaluative tone. Configurations that combine verification with anonymity foster more moderate ratings and reveal internal frictions, while open, public posting environments tilt toward higher positivity and outward-facing praise. In cultural terms, Glassdoor highlights clan and market in pros, whereas Blind surfaces hierarchy and adhocracy in cons, extending research on review polarization beyond valence to the composition of cultural signals (Pavithra and Westbrook, 2022). Methodologically, applying CVF with transformer-based text measures provides a scalable approach to operationalize culture and complements evidence linking clan culture to more favorable employee perceptions (Hartnell et al., 2011).

These results motivate testable propositions about how verification, anonymity, and perceived audience jointly govern voice, and invite work that examines generalizability across domains while connecting cultural signals to outcomes such as turnover and productivity.

\section*{REFERENCES}

\begin{hangparas}{0.25in}{1}
Cameron, K. S., \& Quinn, R. E. (2011). \textit{Diagnosing and changing organizational culture: Based on the competing values framework} (3rd ed.). Jossey-Bass.

Chaudhary, T., Malhotra, P., Mamidi, R., \& Kumaraguru, P. (2023, December). Blind leading the blind: A social-media analysis of the tech industry. In \textit{Proceedings of the 20th International Conference on Natural Language Processing (ICON)} (pp. 470--480). NLP Association of India.

Cloos, J. (2021). Employer review platforms--Do the rating environment and platform design affect the informativeness of reviews? Theory, evidence, and suggestions. \textit{Management Revue, 32}(3), 152--181.

Deng, L., Sun, W., Xu, D., \& Ye, Q. (2021). Impact of anonymity on consumers' online reviews. \textit{Psychology \& Marketing, 38}(12), 2259--2270.

Ding, K., Li, R., Li, Z., \& Hu, S. (2025). Uncovering employee insights: Integrative analysis using structural topic modeling and support vector machines. \textit{Journal of Big Data, 12}(1), 41.

Figini, P., Vici, L., \& Viglia, G. (2020). A comparison of hotel ratings between verified and non-verified online review platforms. \textit{International Journal of Culture, Tourism and Hospitality Research, 14}(2), 157--171.

Hartnell, C. A., Ou, A. Y., \& Kinicki, A. (2011). Organizational culture and organizational effectiveness: A meta-analytic investigation of the competing values framework's theoretical suppositions. \textit{Journal of Applied Psychology, 96}(4), 677--694.

H\"{o}llig, C. (2021). Online employer reviews as a data source: A systematic literature review. \textit{Human Resource Management Review, 31}(2), 100767.

Koch, S., \& Pasch, S. (2023, December). CultureBERT: Measuring corporate culture with transformer-based language models. In \textit{Proceedings of the 2023 IEEE International Conference on Big Data (BigData)} (pp. 3176--3184). IEEE.

K\"{o}nsgen, R., Schaarschmidt, M., Ivens, S., \& Munzel, A. (2018). Finding meaning in contradiction on employee review sites---Effects of discrepant online reviews on job application intentions. \textit{Journal of Interactive Marketing, 43}(1), 165--177.

Mardumyan, A., \& Siret, I. (2023). When review verification does more harm than good: How certified reviews determine customer--brand relationship quality. \textit{Journal of Business Research, 160}, 113752.

Martin-Fuentes, E., Mateu, C., \& Fernandez, C. (2018). Does verifying users influence rankings? Analyzing Booking.com and TripAdvisor. \textit{Tourism Analysis, 23}(1), 1--15.

Mayzlin, D., Dover, Y., \& Chevalier, J. (2014). Promotional reviews: An empirical investigation of online review manipulation. \textit{American Economic Review, 104}(8), 2421--2455.

Pacelli, J., Shi, T., \& Zou, Y. (2022). Communicating corporate culture in labor markets: Evidence from job postings. \textit{Management Science, 68}(11), 8043--8063.

Pavithra, A., \& Westbrook, J. (2022). An assessment of organisational culture in Australian hospitals using employee online reviews. \textit{PLoS ONE, 17}(9), e0274074.

Seo, J., \& Lee, S. (2021). The moderating effect of organizational culture type on the relationship between cultural satisfaction and employee referral intention: Mining employee reviews on Glassdoor.com. \textit{Journal of Organizational Change Management, 34}(5), 1096--1106.

Sockin, J., \& Sojourner, A. (2023). What's the inside scoop? Challenges in the supply and demand for information on employers. \textit{Journal of Labor Economics, 41}(4), 1041--1079.

Winkler, R., \& Fuller, T. (2019, July 25). Companies manipulate Glassdoor by inflating rankings. \textit{The Wall Street Journal}. \url{https://www.wsj.com/articles/companies-manipulate-glassdoor-by-inflating-rankings-and-pressuring-employees-11548171977}
\end{hangparas}

\end{document}